\input epsf
\documentstyle[prb,aps]{revtex}
\begin{document}
\draft
\title{POWER-LAW CORRELATIONS AND ORIENTATIONAL GLASS
IN RANDOM-FIELD HEISENBERG MODELS}
\author{Ronald Fisch}
\address{Department of Physics\\
Washington University\\
St. Louis, Missouri 63130}
\date{Submitted to Physical Review B, 28 July 1997}
\maketitle
\begin{abstract}
Monte Carlo simulations have been used to study a discretized Heisenberg
ferromagnet (FM) in a random field on simple cubic lattices.  The spin
variable on each site is chosen from the twelve [110] directions.  The
random field has infinite strength and a random direction on a fraction $x$
of the sites of the lattice, and is zero on the remaining sites.
For $x=0$ there are two phase transitions.  At low temperature there
is a [110] FM phase, and at intermediate temperature there is a [111] FM
phase.  For $x>0$ there is an intermediate phase between the paramagnet
and the ferromagnet, which is characterized by a $|{\bf k}|^{-3}$ decay of
two-spin correlations, but no true FM order.  The [111] FM
phase becomes unstable at a small value of $x$.  At $x=1/8$ the [110]
FM has disappeared, but the power-law correlated phase survives.
\end{abstract}
\pacs{PACS numbers: 64.70.Pf, 75.10.Nr, 75.40.Mg, 75.50.Lk}
\section{INTRODUCTION}

Orientational glasses have been a subject of significant interest for over
fifteen years.  An extended review of the experiments has been given by
H\"ochli, Knorr and Loidl,\cite{1} and a companion review of the theories has
been provided by Binder and Reger.\cite{2}  More specialized reviews of
electric dipole glasses\cite{3} and the solid hydrogens\cite{4} are also of
interest.  A primary question, which has remained unresolved to the present
time, has been whether the orientational freezing transition which is seen
in the experiments represents a true thermodynamic phase transition, with a
definite critical temperature, $T_c$.  The alternative point of view is
that the freezing should be explained as a kinetic effect, which occurs
gradually, over a range of temperature.  It is surely true that some of the
experiments should be explained as kinetic effects.  The question is
whether a phase transition of this type is possible in a three-dimensional
system with realistic interactions.

It was argued by Michel,\cite{5} who performed a mean-field calculation,
that the proper model for studying the orientational glass is an $n=3$
Heisenberg model, (where $n$ is the number of spin components) with
the addition of a cubic symmetry-breaking term due to the single-ion
anisotropy, and a random-field term due to the alloy disorder.  In this
work we will use a Hamiltonian of the form
\begin{equation}
 H_{RF} = - J\sum_{<ij>} \sum_{\alpha = 1}^3 S_{i}^\alpha S_{j}^\alpha
    - K\sum_i \sum_{\alpha = 1}^3 ( S_{i}^\alpha )^4
    - G\sum_{i^\prime} ({\bf S}_{i^\prime} \cdot{\bf n}_{i^\prime} - 1 )\, ,
\end{equation}
where the sites $i$ form a simple cubic lattice and $<ij>$ indicates a sum
over nearest neighbors.  The $\alpha$ are spin indices, each
${\bf n}_{i^\prime}$ is an independently chosen random unit vector, and the
$i^\prime$ sites are a randomly chosen subset of the lattice containing a
fraction $x$ of the sites.  Since the defect sites are assumed to be
immobile, the random fields do not change with time.  As the random fields
only occur on a fraction $x$ of the sites, Eq.~(1) is a diluted random-field
Heisenberg model.  There are several alternative choices of a Hamiltonian
for an orientational glass,\cite{2} such as using dipolar exchange or
quadrupolar exchange instead of the isotropic exchange chosen here.  In our
present state of knowledge (or ignorance), it may be believed that this
choice does not change the qualitative behavior of the orientational
freezing transition.  We will comment on this point later.

The $G$ term gives the interaction of the orientational order with the point
defects.  In this work we will study the strong-pinning case of Fukuyama and
Lee,\cite{14} in which $G$ is taken to be large, and the density of defect
sites, $x$, is taken to be small.  In contrast to the one-dimensional case
considered by Fukuyama and Lee, the strong-pinning case becomes highly
nontrivial in three dimensions when $x$ is small compared to $1-p_c$,
where $p_c$ is the site percolation concentration.\cite{15}

It has been accepted for some time that, in the absence of some
symmetry-breaking term such as the cubic term, a random field which couples
linearly to the order parameter will always destroy the long-range
symmetry-breaking order of a three-dimensional system which has a continuous
symmetry.  This result was first derived by Larkin,\cite{6} and a simpler
version of the argument was later presented by Imry and Ma.\cite{7}  The
absence of long-range order for systems with short-range interactions in the
presence of random fields in less than four dimensions has been given a
more rigorous basis in the work of Aizenman and Wehr.\cite{8}  The simple
Imry-Ma argument\cite{7} also makes the stronger claim that a weak random
field will cause the ferromagnetic correlation length to be finite at all
temperatures in three dimensions.  This has recently been questioned for the
case of the $XY$ model (i.e. $n=2$) by Gingras and Huse,\cite{9} who argue
that these calculations have not taken proper account of the effects of
vortex loops.  Gingras and Huse suggest that for weak random fields there
may be a phase in which there is quasi-long-range order (QLRO), and the
two-point correlations have a power-law decay as a function of distance.
A recent Monte Carlo calculation by the author\cite{10} has provided strong
evidence in favor of this possibility.

The simple Imry-Ma argument\cite{7} assumes that the low-energy direction
of the magnetization $\bf M$ in some domain is determined by the average
value of the random field over that domain.  While it is surely true that
the ${\bf k}=0$ Fourier component of the random field is the most important
single component in a model with ferromagnetic exchange, it is not clear
that its influence always outweighs the combined effects of all of the
other Fourier components.  The Imry-Ma argument gives the correct lower
critical dimension\cite{8} for the existence of ferromagnetic order in the
presence of a random field.  However, the situation in three dimensions is
very subtle.  When we already know that the ${\bf k}=0$ component of $\bf M$
is not dominant over the other small-wavenumber components, it is no longer
safe to assume that the ${\bf k}=0$ component of the random field is the
only important component of the random field.  Competition between different
small-wavenumber components of the random field may create multiple
low-energy minima of the free energy.

Based on renormalization group calculations, it was suggested by Mukamel and
Grinstein\cite{11} that it might be possible to induce a QLRO phase in an
$n=3$ random-anisotropy system by adding a cubic crystalline anisotropy.
This occurs because a weak cubic anisotropy does not destabilize the $n=3$
Heisenberg critical point in three dimensions, but it removes the Imry-Ma
instability.  This should apply to the random-field case as well.
Even in the presence of a weak random field, we expect a ferromagnetic
phase to exist at low temperatures, but there is no longer a stable
renormalization-group critical fixed point which describes the transition.
Thus, as argued by Mukamel and Grinstein, either the phase transition must
become first order, or else there must be some type of intermediate behavior
between the ferromagnet and the paramagnet (PM).  In the $n=2$ case it was
found\cite{10} that both possibilities may occur, at different values of $x$.

At our current level of understanding we have no reliable analytical method
of predicting whether the QLRO which has been found for the $n=2$ case
should also occur for larger values of $n$.  There was, in fact, another
suggestion, by Mayer and Cowley,\cite{12} that QLRO should occur for $n=2$,
and also for $n=3$ when the cubic term is present.  This argument yields a
line of fixed points with continuously varying critical exponents, rather
than the zero-temperature fixed point of Gingras and Huse.\cite{9}  Both of
these analyses seem to use the existence of dislocation lines in an
essential way.  However, there is no direct evidence in the $n=2$ Monte
Carlo results\cite{10} that dislocations are crucial.  Based on the work of
M\'ezard and Young,\cite{13} one should expect the existence of replica
symmetry breaking in the QLRO phase.

We will find that adding a special type of weak cubic anisotropy does allow
us to find a QLRO phase, and we will study some of its properties.  Because
an orientational glass typically exists in the presence of some crystalline
lattice potential, this result may be directly applicable to cases of
experimental interest.  It should be noted that our model does not allow
the existence of any lattice defects such as dislocation lines or stacking
faults, although these may be needed to explain some of the experimentally
observed behavior.  We will not conclusively answer here the question of
whether the QLRO phase exists also for the isotropic $n=3$ random field
model, but we will present indirect evidence in favor of this.  If it does
exist even in that case, its properties should be essentially those that we
find here.

\section{DISCRETIZED HEISENBERG MODELS}

Rapaport\cite{16} has discussed the technical advantages of using a
discretization of the sphere for Monte Carlo calculations of the Heisenberg
model.  He also suggested that this might be a useful way of studying
the effects of random fields on this model.  There is one significant point
which was not made in his work and which will be important here.  The
discretization of the sphere chosen by Rapaport was the set of 30 unit
vectors defined by the centers of the edges of an icosahedron, which we will
refer to as the ${\bf I}_{30}$ model.  In the work presented here we will
use the set of 12 unit vectors defined by the edge-centers of a cube, which
in the standard Cartesian notation are called the [110] vectors.  We will
refer to this as the ${\bf O}_{12}$ model (where ${\bf O}$ stands for
octahedral).  In either case, with the simple Heisenberg exchange Hamiltonian
\begin{equation}
  H = -J \sum_{<ij>} \sum_{\alpha = 1}^3 S_{i}^\alpha S_{j}^\alpha \, ,
\end{equation}
where each ${\bf S}_i$ is now restricted to the discrete set, rather than
the entire sphere, there are now two phase transitions.  There is the $n=3$
Heisenberg critical point, which occurs at a slightly higher $T_c$
than in the ordinary Heisenberg model, and, in addition, there is a
first-order phase transition at a lower $T_c$, at which the minimum
free-energy directions of $\bf M$ change.

The nature of the low temperature transition can be understood in a Landau
mean-field treatment.  We will discuss the ${\bf O}_{12}$ case here, and
Rapaport's ${\bf I}_{30}$ case is essentially similar.  Because we now have
only a cubic symmetry of the Hamiltonian rather than the full symmetry of
the sphere, the Landau free-energy functional $F(T,{\bf M})$ should contain
terms proportional to $\sum_{\alpha} (M^{\alpha})^{2l}$ for all (positive)
integer values of $l$, in addition to the usual isotropic terms.  Since we
want to find the minimum of the free energy, it is sufficient to consider
only the high-symmetry directions: [100], [110] and [111].

The coefficient of the term of order $M^{2l}$ is proportional to $8(1/2)^l$
for [100], $2+8(1/4)^l$ for [110], and $6(2/3)^l$ for [111].
The $l=1$ term is, of course, isotropic, and the $l=2$ term favors the [111]
directions.  Thus, just below the critical point, where the thermal average
$\langle |{\bf M}| \rangle$ is small, $\langle{\bf M}\rangle$ will be
aligned along [111].

All of the terms with $l>2$ favor the [110] directions.  Therefore, as we go
to still lower $T$, and $\langle |{\bf M}| \rangle$ grows, we eventually
reach a point where it becomes favorable to rotate $\bf M$ to a [110]
direction.  Because there are 8 [111] directions and 12 [110] directions,
and $12/8$ is not an integer, the Landau rules determine that this
transition must be first order.

It is natural to expect that the behavior of the ${\bf I}_{30}$ model would
be closer to the isotropic limit than that of the ${\bf O}_{12}$
discretization used here.  This is surely true at most $T$.  For example,
it is likely that the reason Rapaport did not report the existence of the
low temperature transition is that in the ${\bf I}_{30}$ model it occurs at
a $T_c$ which is smaller than the minimum used in his Monte Carlo
simulations.  However, because the 12 [110] states are not a subset of the
30 icosahedral edge-center states, results for both the ${\bf O}_{12}$ model
and the ${\bf I}_{30}$ model cannot be combined in a simple way to obtain an
extrapolation to the isotropic limit.

One might think at first that a similar intermediate temperature FM phase
would occur for the discretization based on the edge-centers of a
tetrahedron.  These 6 states, however, are equivalent to the 6 [100] vectors
of a cube.  This model has only two phases, with a first-order phase
transition between the FM and PM phases.\cite{17}

Our discretization of the spin variables using the 12 [110] vectors
automatically builds a cubic anisotropy into the free energy.  Thus, since
we are only trying to understand the qualitative aspects of orientational
ordering and not attempting a quantitative model of a particular experiment,
we do not need to keep the $K$ term in the Hamiltonian.  If we now take the
limit $G \to \infty$ with each of the ${\bf n}_{i^\prime}$ chosen from the
set of [110] unit vectors, to simplify the calculation, Eq.~(1) reduces in
form to Eq.~(2), with the dynamical restriction that a fraction $x$ of the
spins (the ones on the $i^{\prime}$ sites) point in fixed random directions,
parallel to the local random fields on the $i^\prime$ sites.

It is interesting to note that the stability of the [111] phase is of purely
entropic origin.  The energy always favors the [110] alignment.
A significant consequence of this is that the walls between different [111]
domains must be broad, rather than sharply defined.  As a result, the [111]
FM phase is easily destabilized by a random field.

\section{MONTE CARLO CALCULATION}

Because all of the ${\bf S}_i$ are chosen from the [110] set, Eq.~(2) has
the helpful property that the energy of every state is an integral multiple
of 1/2.  Thus it becomes easy to write a Monte Carlo program to study
Eq.~(2) which uses integer arithmetic to calculate energies.  This, plus the
fact that that each spin has only 12 possible states, gives substantial
improvements in performance over working with the general form of Eq.~(1),
for both memory size and speed.  It is also be possible to use integer
arithmetic if $G$ is chosen to be an integer.  When $G$ is infinite, the
random field is implemented by assigning a fraction $x$ of the sites to be
the $i^\prime$ sites.  The spins on these sites are given random [110]
directions, and then left fixed for the remainder of the calculation.

The Monte Carlo program used two linear congruential pseudorandom number
generators.  In order to avoid unwanted correlations, the random number
generator used to select which sites would be assigned the random fields
was different from the one used to assign the initial ${\bf S}_i$.  A heat
bath method was used for flipping the spins, which at each step reassigned
the value of a spin to one of the twelve allowed states, weighted according
to their Boltzmann factors and independent of the prior state of the spin.

$L \times L \times L$ simple cubic lattices with periodic boundary
conditions were used throughout.  The values of $L$ used ranged from 16 to
64.  Away from any $T_c$ the samples were run for 10,240 Monte Carlo steps
per spin (MCS) at each $T$, with sampling after each 10 MCS.  Near a $T_c$
they were run several times longer.  The initial part of each data set was
discarded for equilibration.  Typically, two different random field
configurations with a given $L$ were studied for a given $x>0$.  This gives
a rather crude estimate of the finite size dependence of the various
thermodynamic properties.  Unfortunately, however, in the presence of the
random field high precision finite-size scaling is not a very effective
tool, because the sample-to-sample variations for a given size are large
and not well-behaved.\cite{10,18}

Both random and ordered initial conditions were used for the unpinned spins.
For large samples, typically two or three random initial states were tested,
and brief tests were made of all of the 12 ordered states.  For $x=1/16$ (or
less), in a ground state essentially all of the unpinned spins are aligned
along one of the [110] directions.  Thus, in these cases it is easy to
equilibrate the system at low temperatures, by starting from an ordered
state.  This is not true for $x=1/8$, however.

In the absence of any external field, random or uniform, the rotation of
$\bf M$ between different [110] directions is a slow process.  Because all
of these 12 directions are equivalent, however, there is no need for the
Monte Carlo program to average over the minima in this case.  In the
presence of the random field the different [110] ferromagnetic Gibbs states
have different energies.  If the system is started in a high-energy [110]
direction, it will eventually jump to a more favorable direction (unless $T$
is so low that this does not happen in the time available).  In this, as in
many other respects, the model behaves like the random-field Ising
model.\cite{18}

\section{NUMERICAL RESULTS FOR \lowercase{$x = 0$}}

Specific heat data for the pure ($x=0$) ${\bf O}_{12}$ model with $L=32$ and
$L=48$, obtained by numerically differentiating the energy, are displayed
in Fig.~1.  The Heisenberg critical point occurs at
$T_{c1}/J = 1.453 \pm 0.001$, which is less than 1\% above the value found
for the standard isotropic Heisenberg model,\cite{19,20} which is about
$1.4430 \pm 0.0001$.  The $T_c$ for the isotropic model is known to greater
precision because far more computing resources have been used to calculate
it.  As argued by Rapaport,\cite{16} given equal resources one should be
able to get more than equal precision for the discretized model.  Within
the accuracy calculated here, the most effective way of estimating $T_c$
is to assume that the energy $E (T_c)$ is the same for both models.  Our
estimate of $T_{c1}$ for the ${\bf O}_{12}$ model is based on
$E(T_{c1})=0.994$, as found in the isotropic model.\cite{19,20}  It is,
however, necessary to consider the $L$ dependence of $E$.  This method also
works for the $n=2$ case,\cite{10} and reflects the fact that the two-spin
correlation function at the Heisenberg critical point is virtually identical
at all distances in the discretized model and the isotropic model, even
though strict universality only demands that they be identical at large
distances.  The short distance part of the two-spin correlation function,
and thus $E (T_c)$, does depend on the lattice structure, even in a
Bethe-Peierls mean-field treatment.

The first-order transition from the [111] FM phase to the [110] FM phase is
indicated in Fig.~2 by the vertical arrow.  This transition occurs at
$T_{c2}/J=1.0625 \pm 0.0075$, where the error bar indicates the approximate
width of the region of metastability, rather than a statistical error.
Measured at $T/J=1.0625$, the latent heat at this transition is measured
to be $\Delta Q=0.0561~J$ for $L=32$, and $\Delta Q=0.0571~J$ for $L=48$.
The observed increase of $\Delta Q$ as $L$ increases confirms that the
transition is indeed first order.  Because the specific heat is
substantially larger in the [110] phase than in the [111] phase near
$T_{c2}$, the largest contribution to the uncertainty in $\Delta Q$ comes
from the uncertainty in $T_{c2}$.

The magnetization $\langle |{\bf M}(T)| \rangle$ is shown in Fig.~2, for
$L=24$, 32 and 48.  We see that the size dependence rapidly becomes small
below $T/J=1.40$.  The jump in $|{\bf M}|$ at the first-order transition,
again using data taken at $T=1.0625$, is 0.0277 for $L=32$ and 0.0286 for
$L=48$.  The finite-size scaling function for $|{\bf M}(T)|$ at the
Heisenberg critical point is displayed in Fig.~3.  The values of the
critical exponents used for this figure are the usual field theory
estimates\cite{21} for $n=3$.  Thus, if we determine $T_c$ by the energy
condition discussed above, there are no free fitting parameters.  The width
of the scaling region is slightly larger than that of another $n=3$
model\cite{22} for which the cubic anisotropy is probably somewhat stronger.

\section{NUMERICAL RESULTS FOR \lowercase{$x > 0$}}

In addition to $x=0$, Monte Carlo data were also obtained at $x=1/32$, 1/16,
and 1/8.  A semi-quantitative picture of the phase diagram obtained from
these results is shown in Fig.~4.  The limit of stability of the [110] FM
ground state is slightly less than $x=1/8$.  For the nonzero values of $x$
used in these calculations, the [110] FM has a transition into the QLRO
phase.  The [111] FM phase, which is stable at $x=0$, should also extend to
small positive values of $x$.  As noted above, however, the domain walls in
this phase are broad and have a low cost in free energy.  Thus, the [111]
FM is easily destabilized by the random field.  It is difficult to obtain
meaningful numerical results for very dilute random fields, due to crossover
effects.\cite{23}  Therefore, the [111] FM-QLRO phase boundary was not
observed directly, and its existence is shown in Fig.~4 as a dotted line.

The QLRO-to-PM transition is second order for small $x$ and first order for
larger $x$, with a tricritical point somewhere between $x=1/16$ and $x=1/8$.
The shift in $T_c$ for the QLRO-to-PM transition is linear in $x$ for small
$x$, with a slope of
\begin{equation}
 {d \over dx}\biggl({{T_c (x)} \over {T_c (0)}}\biggr)=-3.75\pm 0.20 \quad .
\end{equation}
This should be compared to the corresponding quantity for the $n=2$ case,
which is\cite{10} $-3.42 \pm 0.14$.  ($x=1/16$ is in the linear regime for
$n=2$ also.)  Having the shift in $T_c (x)$ increase by about 10\% as one
goes from the $n=2$ case to the $n=3$ case is about what should be expected
if the QLRO phase survives for isotropic spin variables.  If the existence
of the QLRO was actually dependent on the discretization, the author would
expect this shift to be substantially larger for the ${\bf O}_{12}$ model
than we have found it to be. 

It is interesting to note that these numbers are about three times as large
as the shifts in $T_c (x)$ which one obtains from a simple quenched site
dilution,\cite{24} although a naive mean-field approximation predicts that
the shift should be the same in the two cases.  It is also true in the
site dilution case that the size of the shift increases as $n$ increases.

Our results indicate that the QLRO phase still occurs for $x=1/8$ when $n=3$,
while for $n=2$ it was found that the limit of stability of the QLRO phase
was\cite{10} less than $x=1/8$.  This is an indication that the stability
of the QLRO phase increases with $n$.  On this basis, we might expect it to
occur for all $n$, and thus its existence would be independent of any
topological singularities.  This encourages us to hope that it might be
possible to study the QLRO using some analytical technique which uses $1/n$
as an expansion parameter.  The existence of such an expansion must, for now,
be considered mere speculation.  It could certainly still be true that the
QLRO phase really is of topological origin, in which case it should not exist
for $n>3$.

It is difficult to study transitions which occur at low $T$ using
Monte Carlo calculations.  The $T=0$ endpoints of both the [110] FM-QLRO 
boundary and the QLRO-PM boundary in Fig.~4 are illustrated in schematic
fashion.  The author does not mean to imply that $T_c (x)$ is actually
linear in $x$ near the $T=0$ endpoint in either case.

The evolution of the specific heat as $x$ is increased is shown in Fig.~5.
The data displayed were obtained by numerically differentiating the
calculated values of the energy with respect to $T$.  The specific heat was
also computed by calculating the fluctuations in the energy at fixed
temperature, yielding similar but noisier results.  We see that the data
for different samples with the same value of $x$ agree fairly well, although
some differences are visible near the phase transitions.

The sharp peaks which occur for $x=0$ (in Fig.~1) have become rounded at
$x=32$, and they have moved to lower $T$.  The QLRO-to-PM transition
actually occurs slightly below the $T$ of the upper specific heat peak.
The [110] FM-to-QLRO transition appears to be continuous, rather than
first order, and the specific heat increases as we approach this transition
from either direction.  Going on to $x=1/16$, all of these trends are
enhanced.  At $x=1/8$, the FM phase has disappeared entirely.  Due to the
long relaxation times, for $x=1/8$ it was not practical to equilibrate an
$L=64$ lattice at $T/J<0.65$.  The small specific heat peak near $T=0.67$
marks the QLRO-to-PM transition, which is first order at this value of $x$.
For $x=1/32$ and $x=1/16$, where this transition is second order, there is
no clear signature of the transition observable in the specific heat.

Looking at the dependence of $\langle |{\bf M}| \rangle$ on $x$ and $L$
provides additional insight.  The data for $x=1/32$, 1/16, and 1/8 are shown
in Fig.~6.  In the [110] FM phase, $\langle |{\bf M}| \rangle$ is almost
independent of $L$, except very close to $T_c$.  In the QLRO phase,
$\langle |{\bf M}| \rangle$ decreases slowly as $L$ increases, probably
decaying as $1/ \log (L)$.  In the PM phase, $\langle |{\bf M}| \rangle$
decreases as $(L/\xi)^{-3/2}$, where $\xi$ is the ferromagnetic correlation
length.  In Fig.~6c we see that the data for large $L$ and $T$ between 0.70
and 1.45 seem to form a set of parallel lines on this semi-log plot.  This
implies that $\xi$ remains constant over this range of temperature for
$x=1/8$.  The saturation of $\xi$ in the PM phase indicates that the
QLRO-to-PM transition is first order at this value of $x$.

We can get valuable information by looking at the magnetic structure
factor of samples of size $L=64$.  The structure factor is the spatial
Fourier transform of $\langle {\bf M}^2\rangle$, and it can be measured by
X-ray and neutron scattering experiments.  Near a critical point the
small-wavenumber behavior of the structure factor of a random-field model
is expected to have the form
\begin{equation}
          | \langle {\bf M} ({\bf k}) \rangle |^2 
  \approx ( 1/\xi^2 + |{\bf k}|^2)^{-(4-\bar\eta )/2} \quad .
\end{equation}
The correlation length $\xi$ is infinite in the QLRO phase (essentially by
definition).

In three dimensions, Eq.~(4) requires that $\bar\eta \geq 1$.  To estimate
$\bar\eta$, we measure the slope of the structure factor on a log-log plot.
This is shown, averaged over angles, for two $L=64$ lattices with $x=1/16$
at $T=1.125$, in Fig.~7a.  The slope of the best fit to the data at small
$|{\bf k}|$ is $-2.87\pm 0.05$, so we find
\begin{equation}
    \bar\eta = 1.13\pm 0.05 \quad .
\end{equation}
This is essentially the same as the result found for this quantity in the
$n=2$ case.\cite{10}  It is, however, well known that the value of an $\eta$
exponent is often insensitive to the value of $n$, so one should not believe
on these grounds that the values of $\bar\eta$ for $n=2$ and $n=3$ are
identical.  The data for $x=1/32$ (not shown) have a smaller value of the
slope at $T_c$.  This, however, should not be interpreted as measuring the
value of $\bar\eta$, because these data are taken in the crossover
region\cite{23} from pure system behavior, which has a slope of $-1.97$ for
the structure factor at $T_c$.

Repeating the above procedure for two $L=64$ with $x=1/8$ lattices at
$T=0.6875$, using cold start initial conditions, we find the results shown
in Fig.~7b.  This value of $T$ is slightly above the best estimate of $T_c$
at this value of $x$, and the measured values of $\langle |{\bf M}| \rangle$
for these data are 0.2518 for $S1$ and 0.2880 for $S2$.  Similar states, but
with slightly higher energy and lower magnetization, were achieved using
random initial states.  We see that the structure factor again shows a
power-law behavior at small $|{\bf k}|$, but that the slope at $T_c$ has now
assumed the minimum allowed value of $-3$.  This is the value inside the
QLRO phase, and yields
\begin{equation}
    \bar\eta_0 = 1  \quad .
\end{equation}
This is an independent confirmation of the first-order nature of the
transition at this value of $x$.  The latent heat is too small to measure
accurately, being less than $0.01~J$.  The small latent heat may be an
indication that we are close to the tricritical point.

Although no first-order behavior was seen along the QLRO-to-PM transition
line\cite{10} for $n=2$, it probably exists for $x$ near the endpoint of the
QLRO phase in that case also.  The tricritical point on the QLRO-to-PM
transition line is likely to have the same origin as the analogous
tricritical point which occurs for the random-field Ising model.\cite{25}

\section{DISCUSSION}

What we have found is that, despite the weak destruction of the
ferromagnetic long-range order caused by the Imry-Ma instability, there is
actually a good correspondence in three dimensions between the behavior of
the QLRO-to-PM transition in $n=2$ and $n=3$ random-field models and the
FM-to-PM transition in the random-field Ising model.  The author sees no
reason why the QLRO phase should become completely destabilized by a weak
uniaxial anisotropy.   Therefore, it should be possible for an $n=3$ system
with a weak uniaxial anisotropy and a random field to exhibit a QLRO phase
between its PM and FM phases.  This is, in fact, a good description of the
diluted antiferromagnet in a field systems which serve as the prototypical
experimental random-field Ising models.  Therefore, the ``metastability''
which is seen in these experimental systems\cite{26} is probably caused by
the presence of a QLRO phase, at least those cases where the uniaxial
anisotropy is weak.

Since the QLRO phase exists for both $n=2$ and $n=3$ in three dimensions,
and possibly for larger values of $n$ also, there cannot be any simple
correspondence between the random-field models and the pure $n$-vector
models in one lower spatial dimension.  We do not yet know if the QLRO phase
exists for $n \ge 4$, so we cannot say whether the QLRO is properly
attributed to topological excitations, as the Kosterlitz-Thouless phase in
the pure model is.

As mentioned in the Introduction, many orientational glass systems have
quadrupolar interactions, rather than the vector interactions studied here.
In cases where the parallel alignment of the quadrupoles is the low-energy
one, such as the diluted alkali-cyanides,\cite{1} our model should apply,
at least to the extent that lattice defects can be neglected.
In those cases where the low-energy alignment is the T-configuration,
however, such as the solid hydrogens,\cite{4} things are more complicated.
Another interesting experimental system with a quadrupolar order parameter,
for which our results should be relevant, is the isotropic-nematic
transition of a liquid crystal in silica gel.\cite{27}

\section{CONCLUSION}

In this work we have used Monte Carlo simulations to study the ${\bf O}_{12}$
version of the diluted random-field ferromagnet in three dimensions.
We have found that there are two types of ordered phases, just as in the
$n=2$ case.  In addition to the anisotropy-stabilized ferromagnet, we find
an intermediate phase displaying a $|{\bf k}|^{-3}$ decay of two-spin
correlations.  There is a tricritical point on the QLRO-to-PM transition
line.  When this transition is second order, the critical exponent
$\bar\eta$, which characterizes the magnetic structure factor on the
critical line, has a value which is indistinguishable from its value in
the $n=2$ case.  The exponent $\bar\eta_0$, which is observed within the
QLRO phase, is also the same as in the $n=2$ case.  The results should be
applicable to a variety of experimental systems.

\acknowledgments

The author is greatful to Michael Aizenman, Brooks Harris, Frances Hellman,
David Huse, David Landau and Phil Taylor for helpful discussions during the
course of this work.

\newpage
\begin{figure}
\caption{
Specific heat vs. temperature for the pure ${\bf O}_{12}$ model
on $L \times L \times L$ simple cubic lattices.  The large vertical arrow
indicates a first-order phase transition.}
\label{fig1}
\end{figure}

\begin{figure}
\caption{
Magnetization vs. temperature for the pure ${\bf O}_{12}$ model
on $L \times L \times L$ simple cubic lattices.}
\label{fig2}
\end{figure}

\begin{figure}
\caption{
Finite-size scaling of the magnetization near $T_c$ for the pure
${\bf O}_{12}$ model on $L \times L \times L$ simple cubic lattices.
The $y$-axis is scaled logarithmically.}
\label{fig3}
\end{figure}

\begin{figure}
\caption{
Phase diagram of the dilute random-field ${\bf O}_{12}$ model on simple
cubic lattices, showing the paramagnetic (PM), ferromagnetic (FM), and
quasi-long-range order (QLRO) phases.  The plotting symbols show estimates
obtained from the Monte Carlo data.  The solid lines indicate first-order
transitions, the dashed lines indicate second-order transitions, and there
were no data taken on the dotted line.}
\label{fig4}
\end{figure}

\begin{figure}
\caption{
Specific heat vs. temperature for the dilute random-field ${\bf O}_{12}$
model on $L \times L \times L$ simple cubic lattices.
(a)~$x=1/32$; (b)~$x=1/16$; (c)~$x=1/8$.}
\label{fig5}
\end{figure}

\begin{figure}
\caption{
Magnetization vs. temperature for the dilute random-field ${\bf O}_{12}$
model on $L \times L \times L$ simple cubic lattices.  The $y$-axis is
scaled logarithmically. (a)~$x=1/32$; (b)~$x=1/16$; (c)~$x=1/8$.}
\label{fig6}
\end{figure}

\begin{figure}
\caption{
Angle-averaged magnetic structure factor for the dilute random-field
${\bf O}_{12}$ model on $64 \times 64 \times 64$ simple cubic lattices,
log-log plot.  Each data set shows averaged data from 2 states sampled at
10,240 MCS intervals.
(a)~$x=1/16, $$T=1.125$, the line has a slope of $-2.87$;
(b)~$x=1/8$, $T=0.6875$, the line has a slope of $-3.00$.  Note that the
vertical scales differ in (a) and (b).}
\label{fig7}
\end{figure}

\end{document}